\RequirePackage{lineno}
\documentclass[aps,prc,twocolumn,nofootinbib,superscriptaddress, 
]{revtex4}

\usepackage{epsfig}
\usepackage{amssymb}
\usepackage{bm}
\usepackage{graphicx,color}
\usepackage{dcolumn}
\usepackage{epstopdf}

\usepackage{ulem}

\renewcommand{\emph}[1]{\textit{#1}}

\newcommand{\eftnopi}{\mbox{EFT($\not \! \pi$)}}

\begin{document}

\title{Parity Violation in Photonuclear Reactions at HIGS\\
Submission to Snowmass 2013: Intensity Frontier}
\author{M.W. Ahmed}
\affiliation{Department of Mathematics and Physics, NC-Central Univ., Durham, NC}
\author{A.E. Champagne}
\affiliation{University of North Carolina, Chapel Hill, NC}
\author{B.R. Holstein}
\affiliation{University of Massachusetts, 
Amherst, MA 
}
\author{C.R. Howell}
\affiliation{Department of Physics, Duke University, Durham, NC}
\author{W.M. Snow}
\affiliation{Indiana University/CEEM, 
Bloomington, IN 
}
\author{R.P. Springer}
\affiliation{Department of Physics, Duke University, Durham, NC}
\author{Y.K. Wu}
\affiliation{Department of Physics, Duke University, Durham, NC}
\date{\today}

\begin{abstract}
We discuss the scientific motivation, possible experiments,  and beam requirements for measurements of parity violation in photonuclear reactions at an intensity upgraded HIGS facility, HIGS2.  
\end{abstract}


\maketitle

HIGS2 PV Working Group members:  
M. Ahmed, Duke; 
J.-W. Chen, National Taiwan U.;
C. Crawford, University of Kentucky;
W. Deconinck, William and Mary;
N. Fomin, LANL;
H. Gao, Duke;
M. Gericke, U Manitoba;
L. Girlanda, INFN;
H. Griesshammer, GWU;
V. Gudkov, U South Carolina;
S. Jawalker, Duke;
H. Hammer, Bonn;
B. Holstein, University of Massachusetts;
C. Howell, Duke;
P. Huffman, NC-State;
C.H.Hyun, Daegu;
S. Kucuker, UT-Knoxville;
D. Lee, NC-State;
C.-P. Liu, NDHU;
D. Markoff, NCCU;
B. Norum, UVA;
M. Schindler, U South Carolina;
P. Seo, University of Virginia;
W. M. Snow,  Indiana University/CEEM;
R. P. Springer, Duke University;
J. Vanasse, U Mass;
S. Wilburn, LANL;
B. Wojtsekhowski, JLAB;
Y.K. Wu, Duke;
HangHua Xu, Shanghai Institute of Applied Physics;
Wang Xu,  Shanghai Institute of Applied Physics;
Shi-Lin Zhu, Peking U;

\section{Introduction}

This submission to Snowmass 2013 discusses the possibility of performing photonuclear reactions at an upgraded HIGS facility (HIGS2) that can observe parity violation (PV) induced from the weak interaction between nucleons. 
The physics considered here addresses Performance Measure FI8,
``Perform independent measurements of parity violation
in few-body systems to constrain the non-leptonic weak
interaction'' from the 2007 NSAC Performance Measures document.  

At the present time, it is unclear whether currently available measurements involving low energy nucleons are consistent with the Standard Model,
or indeed with each other.   The reason for this is that many such measurements involve heavy nuclei whose strong interactions, binding mechanism, etc., are not understood in terms of QCD. While it is likely that the PV  observables in  heavy nuclei are the result of weak interactions among only two or three constituent nucleons, that physics is not presently extractable from the complicated strong interaction physics involved.  

To unambiguously extract the weak interactions among nucleons requires PV measurements in very light nuclei: the deuteron, tritium, He-3, and now perhaps even up to four nucleons, because the strong interactions in these very light systems are understood in terms of model-independent effective field theories (EFTs) that systematically incorporate the
symmetries of QCD in a consistent fashion.  At leading order, and at very low energies (photon energies below 10 MeV), there are five low-energy PV constants (LECs) that parameterize the physics \cite{Dan,Zhu05,Girlanda:2008ts}.  Before
we know whether the Standard Model as encoded in the relevant EFT, \eftnopi, is correct, all five will have to be determined.  


A complementary point of view is that utilizing weak interactions in few-nucleon systems provides
a unique probe of QCD in these systems, as the observables in question come from interference between weak and strong effects. Another important feature of this
weak probe of low-energy QCD is that it predominantly samples
the two-nucleon interaction at distances
where the volumes of the interacting nucleons overlap, i.e., at
distances less than about 2 fm.

One constraint is available from the low energy (13.6 MeV)  PV longitudinal asymmetry from polarized protons scattering off unpolarized protons \cite{Potter74, Eversheim91},
$A_L = (\sigma_+ - \sigma_- ) /( \sigma_+ + \sigma_-)$ where $\sigma_\pm$ indicates the cross section from protons polarized along/against the incoming proton's momentum.   A second independent measurement is underway at the Oak Ridge National Laboratory SNS, NPDGamma \cite{Ger11}, an experiment that will
measure the angular distribution of the exiting photon after polarized neutron capture on a proton,
$\frac{1}{\Gamma}\,\frac{d\Gamma}{d\cos \theta}=1+A_\gamma \cos \theta,  $ where $A_\gamma$ is the PV-dependent quantity and $\theta$ is the angle between the direction of polarization of the neutron and the photon momentum.   We discuss in this submission to Snowmass the possibility of a third independent measurement: $\vec \gamma d \rightarrow n p$.  Note that this is {\sl not} the same as the time-reversed reaction of the NPDGamma measurement; instead we propose to measure a completely orthogonal observable.  Limits from previous attempts exist from Chalk River~\cite{Earle88}, and the reverse reaction~\cite{Knyaz'kov:1984zz}. Experiments to measure this observable have been proposed at JLAB \cite{Sinclair00,Woj00}, SPRING-8, and
the Shanghai Synchroton.  

What we propose here is an intensity upgrade of the High Intensity Gamma Source (HIGS) at the Triangle Universities Nuclear Laboratory (TUNL).  HIGS2 would provide unprecedented photon luminosity, polarization control, and energy 
resolution.   
Based on the high-current electron storage ring in the Duke Free Electron Laser Laboratory and using known accelerator and laser technology, the expected
capabilities of the HIGS2 are:  $E_\gamma$ = 2 -- 12 MeV
with total gamma-ray
flux of $10^{11}$ to $10^{12}$ photons/second and minimum beam energy resolution
less than 0.5\%; high linear or circular polarization (90\% to 95\%); and rapid
switching of beam polarization.
\baselineskip 11pt
\section{PV in circularly polarized photon breakup of  the deuteron}

Naive dimensional analysis along with estimates of the weak couplings from a mixture of model calculations and existing experiments suggests that the PV asymmetry involved in $\vec \gamma d \rightarrow
 n p$ is on the order of $5 \times 10^{-8}$.  
The relation between this asymmetry and weak couplings has been calculated by several authors, both as a function of gamma energy in hybrid calculations \cite{Lee78, Oka83, Sch04, Fuj04, Liu04, Hyun05} and at threshold, where it is related by time reversal symmetry to the circular polarization of the photon in unpolarized neutron-proton capture~\cite{Schindler:2009wd,Hyun09}. 

The desired statistical uncertainty of $10^{-8}$ requires HIGS2 to supply sufficient 2.3 MeV circularly polarized gamma-rays to
produce $10^{16}$
photodisintegration reactions in a
liquid deuterium 
target 
over a calendar year of operation with 
photon helicity reversal rate of 1 to 100 Hz with negligible
photon helicity-dependent phase space variations.
The photon beam source for HIGS2 will be a high-power Fabry-Perot
cavity driven by a high rep-rate, mode-locked infrared laser,
as the result of recent advancement in the laser technology
in several laboratories including JILA (see, e.g. Ref.~\cite{pupeza}).
Both moderated neutrons from the photodisintegration reaction and gammas
scattered from the target would be monitored in current mode.
The current-mode gamma detectors could be very similar to those presently being used in the NPDGamma experiment, which are demonstrated to be free of systematic errors at the 1 ppb level.
Because of the intense
gamma-ray field around the target due to
Compton scattering from the target, the neutron detector must be insensitive
to gamma-rays. $^{3}$He ion chambers
would be the technology of choice because of their low sensitivity to gamma-rays
relative to neutrons. The gamma-ray
background in the $^{3}$He counters would be determined using chambers filled
with $^4$He gas, which has almost
identical photon interaction probabilities as $^{3}$He[19].  The signal could be formed as the helicity dependence of the ratio of the neutron to gamma signal. A transmission detector for the gammas through the target can be used to monitor some of the gamma beam properties. 

\section{Status}

We are not aware of any other facility in the world that has the potential to reach the
desired photon luminosity and energy resolution as described here for HIGS2.  
The elements of the HIGS2 involving the electron beam source,
the  Fabry-Perot cavity, and the drive laser utilize
 demonstrated technology. A gamma intensity upgrade is
 expected in
 the near future using a more powerful 2 micron drive laser (from 1
 to
 10 W) as the result of anticipated advancement in the laser
 technology.
Even if additional experiments
with slow neutrons are conducted after the NPDGamma experiment, the full complement of leading
order low-energy PV LECs cannot be obtained without measurements involving photons. 
In order to resolve the question of whether we understand PV in nuclei depends upon
our understanding of PV in few-nucleon  systems, which requires the measurements of
$\vec \gamma d \rightarrow np$.

Developing the capabilities and expertise to measure the PV asymmetry in
$\vec \gamma d \rightarrow n p$ to the desired precision
of 10$^{-8}$ will create opportunities for measuring photo-induced PV
asymmetries on other few-nucleon targets,
e.g., the triton or $^3$He.  Such measurements would provide more independent
determinations of the low-energy
PV LECs.

An energy upgrade to HIGS2 in the future may allow access to $^4$He, for example, but in that case the
experiment will need to be analyzed using a pionful EFT, due to the significant -- 28 MeV -- binding energy.   Heavier nuclei, where PV asymmetry
enhancements may exist, are also possible targets.  

In addition to PV measurements, HIGS2 would create opportunities for
measuring photo-induced nuclear reactions
at energies relevant to nuclear astrophysics with sensitivities
substantially higher than can be achieved at currently
operating facilities.  For example, the beam intensity and features at
HIGS2 will enable measurements of inverse
charged-particle capture reactions with unprecedented sensitivity, which in
many cases have significant advantages
over traditional approaches. A central theme of nuclear astrophysics at
HIGS2 would be the structure, evolution and nucleosynthesis of evolved stars
that give rise to supernova explosions. In particular, a main push in this
area at HIGS2 would be
measurements of the $^{12} {\rm C} (\alpha,\gamma)^{16}$O reaction 
down to energies sufficiently low
to unambiguously
resolve the electromagnetic multipole interference problem using the inverse
reaction, $^{16}{\rm O}(\gamma,\alpha)^{12}$C.
This measurement would address DOE Office of Nuclear Physics Performance
Milestone NA4, ``Reduce uncertainties of the most crucial stellar evolution
nuclear reactions (e.g.  $^{12} {\rm C} (\alpha,\gamma)^{16}$O) 
by a factor of two, 
and others (e.g.
the MgAl cycle) to limits imposed by accelerators and detectors.''

An international workshop on the HIGS2 possibility was held at Duke
University June 3--4, 2013.  The outcome was a commitment on the part of
the following scientists to address the specified issues:  (1) beam: Y.K Wu (Duke) and J. Ye (JILA); (2) polarized interactions: B. Norum (UVA); (3) target/detectors: M. Snow (Indiana), D. Markoff (NCCU), P. Huffman (NC-State), and C. Howell (Duke); (4) GEANT simulations: 
UVA and 
SINAP.

\end{document}